\documentclass[PAPER, atlasdraft=false, UKenglish, cernpreprint, texmf, orcidlogo]{atlasdoc}
\usepackage{atlaspackage}
\usepackage{atlasbiblatex}

\usepackage{atlasphysics}
\graphicspath{{logos/}{figures/}}

\usepackage{ANA-TOPQ-2023-22-PAPER-defs}
\usepackage{fixme}
\usepackage{multirow}
\usepackage{multicol}
\fxsetup{status=draft} %
\AtlasTitle{Constraint on the total width of the Higgs boson from Higgs boson and four-top-quark measurements in $pp$ collisions at $\sqrt{s}$~=~13~TeV with the ATLAS detector}

\AtlasAbstract{%
This Letter presents a constraint on the total width of the Higgs boson ($\Gamma_H$) using a combined measurement of on-shell Higgs boson production and the production of four top quarks, which involves contributions from off-shell Higgs boson-mediated processes. This method relies on the assumption that the tree-level Higgs-top Yukawa coupling strength is the same for on-shell and off-shell Higgs boson production processes, thereby avoiding any assumptions about the relationship between on-shell and off-shell gluon fusion Higgs production rates, which were central to previous measurements. The result is based on up to 140~fb$^{-1}$ of proton--proton collisions at a centre-of-mass energy of $\sqrt{s}$~=~13~TeV collected with the ATLAS detector at the Large Hadron Collider. The observed (expected) 95\% confidence level upper limit on $\Gamma_H$ is 450~MeV (75~MeV). Additionally, considering the constraint on the Higgs-top Yukawa coupling from loop-induced Higgs boson production and decay processes further yields an observed (expected) upper limit of 160~MeV (55~MeV).}

\AtlasRefCode{TOPQ-2023-22}

\PreprintIdNumber{CERN-EP-2024-190}

\arXivId{2407.10631}

\HepDataRecord{https://www.hepdata.net/record/156837}

\AtlasJournalRef{\PLB 861 (2025) 139277}
\AtlasDOI{10.1016/j.physletb.2025.139277}

\AtlasCoverEgroupAnalysisTeam{atlas-topq-2023-22-analysis-team@cern.ch}

\hypersetup{pdftitle={ATLAS document},pdfauthor={The ATLAS Collaboration}}

\begin{document}

\maketitle

\section{Introduction}

The discovery of the Higgs boson~\cite{Englert:1964et,Higgs:1964ia,Higgs:1964pj,Higgs:1966ev,Guralnik:1964eu, Kibble:1967sv} by the ATLAS~\cite{HIGG-2012-27} and CMS~\cite{CMS-HIG-12-028} collaborations at the Large Hadron Collider (LHC) has ushered in a new era for particle physics, marked by precision measurements in the Higgs sector. The total width of the Higgs boson (\gammah) is sensitive to the potential presence of beyond the Standard Model (SM) Higgs boson decays that are not covered by direct experimental searches, making it a crucial parameter for exploring new phenomena.
The Higgs boson mass has been measured to be $125.09 \pm 0.24$~\GeV based on LHC Run~1 data collected by the ATLAS and CMS experiments~\cite{HIGG-2014-14}, and more recently to be $125.11 \pm 0.11$~\GeV based on Run~1 and Run~2 data collected by the ATLAS experiment~\cite{HIGG-2022-20}.
At the measured Higgs boson mass value around 125~\GeV, the SM predicts the value of \gammah to be only 4.1~\MeV~\cite{deFlorian:2016spz}.
Due to limited detector resolution, the total width of the Higgs boson is inaccessible via direct measurement of the Higgs boson line shape or flight distance~\cite{CMS-HIG-14-036} at the LHC experiments. However, a combined measurement of on-shell and off-shell Higgs boson production processes can constrain the total width of the Higgs boson based on specific model assumptions~\cite{Kauer:2012hd,Caola:2013yja,Campbell:2013una,Campbell:2013wga}. The on-shell Higgs boson production and decay rates depend on both the Higgs boson coupling strength to particles involved in the production and decay processes, and the total width of the Higgs boson.
In contrast, the off-shell Higgs boson production and decay depend only on the off-shell Higgs coupling strength. If the relation between the on-shell and off-shell Higgs boson couplings is known, the combined measurement of on-shell and off-shell Higgs boson production processes allows the determination of \gammah.

Recently, both the ATLAS and CMS experiments placed constraints on \gammah\ using combined measurements of on-shell and off-shell Higgs boson production processes in the \(ZZ^{(*)}\) final states, yielding 95\% confidence level (CL) upper limits of 10.2~\MeV~\cite{HIGG-2018-32} and 8.5~\MeV~\cite{CMS-HIG-21-013}, respectively, superseding earlier constraints obtained by the two experiments with Run~1 data using the same assumptions~\cite{HIGG-2014-10,CMS-HIG-14-032}.
These measurements assume that the strength modifiers, which are multiplicative factors of the Higgs boson couplings, are the same between off-shell and on-shell production processes specifically for its interactions with gluons and vector bosons. However, the loop-induced effective Higgs-gluon coupling could vary differently between on-shell and off-shell production processes if not-yet-detected coloured beyond SM (BSM) particles are contributing to the Higgs-gluon coupling~\cite{Buschmann:2014sia,Englert:2014aca,Cao:2021kww}.

\begin{figure}[b]
\centering
\subfloat[]{\includegraphics[width=.3\textwidth]{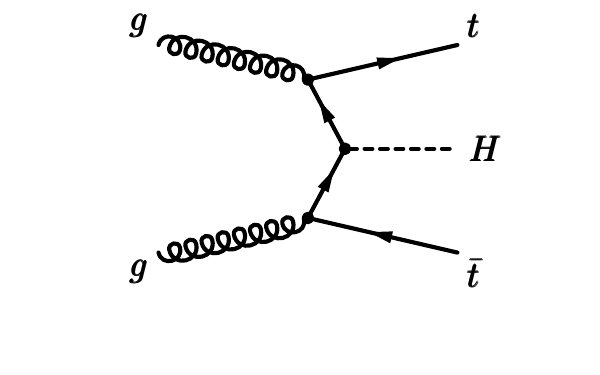}}
\subfloat[]{\includegraphics[width=.3\textwidth]{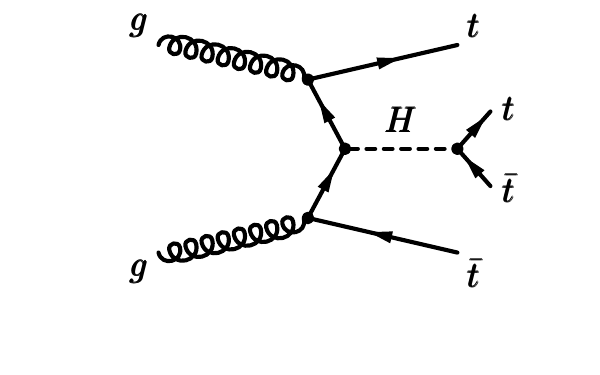}}
\caption{Example leading-order Feynman diagrams for the (a) \ttH and (b) Higgs-mediated \tttt\ processes.}
\label{fig:feyn}
\end{figure}

This Letter presents a constraint on \gammah using a combined measurement of on-shell Higgs boson production in association with a top-quark pair (\ttH) and the simultaneous production of four top quarks (\tttt), which involves contributions from off-shell Higgs boson-mediated processes through the Higgs-top Yukawa coupling. Figure~\ref{fig:feyn} shows representative Feynman diagrams of those processes involving a Higgs boson in the final state or a virtual Higgs boson mediator.
This measurement assumes that the tree-level Higgs-top Yukawa coupling strength is the same for on-shell and off-shell Higgs boson production processes~\cite{Cao:2016wib,Cao:2019ygh}. Different from the Higgs-gluon coupling, the presence of unknown coloured particles would not modify the tree-level Higgs-top Yukawa coupling and its scale dependence. Another assumption made in this study is that no BSM contributions affect the \tttt production. Both ATLAS and CMS experiments have observed \tttt\ production in proton--proton ($pp$) collisions at $\sqrt{s}$~=~13~\TeV and used the measurements to set constraints on possible BSM contributions that may affect this process~\cite{TOPQ-2021-08,CMS-TOP-22-013}. In summary, the constraint on \gammah derived from the combined measurement of the \ttH and \tttt processes represents a complementary approach based on different assumptions from existing studies~\cite{CMS-HIG-21-013,HIGG-2018-32}.

\section{Input measurements}
This combined measurement uses results from individual measurements performed by the ATLAS experiment~\cite{PERF-2007-01} as input, summarised in Table~\ref{tab:inputs}. The sensitivity to the total width of the Higgs boson primarily comes from the \ttH\ and \tttt\ processes. However, measurements targeting other Higgs boson production modes are also included to constrain contributions from Higgs boson couplings beyond its top Yukawa coupling. These additional modes include gluon--gluon fusion (\ggF), vector boson fusion (\VBF), associated production with a vector boson (\WH/\ZH), and associated production with a single top quark (\tH), which also contributes to the Higgs-top Yukawa coupling measurement.

The input measurements rely on the two-level trigger system used to select the events~\cite{TRIG-2016-01}, and a software suite~\cite{SOFT-2010-01} that is used in data simulation, in the reconstruction and analysis of real and simulated data, in detector operations, and in the trigger and data acquisition systems of the experiment.

\begin{table}[t]
\caption{Summary of on-shell and off-shell measurements used as input for the total width measurement. For the on-shell analyses, this table specifically breaks down the processes targeted by the measurement into production and decay modes. The off-shell measurement is included but not broken down in this manner.}
\centering
\begin{tabular}{ccc}
\toprule
\midrule
\multicolumn{2}{c}{Target processes} & Reference \\
\midrule
\multicolumn{3}{l}{Off-shell measurement} \\
\multicolumn{2}{c}{\( pp \rightarrow \tttt \)} & \cite{TOPQ-2021-08} \\
\midrule
\multicolumn{3}{l}{On-shell measurement} \\
\multicolumn{1}{c}{Production} & \multicolumn{1}{c}{Decay} & \\

\ggF, \VBF, \WH, \ZH, \ttH, \tH & \Hyy & \cite{HIGG-2020-16} \\
\ttH + \tH & \Hbb & \cite{HIGG-2020-23} \\
\WH, \ZH & \Hbb & \cite{HIGG-2018-51, HIGG-2018-52} \\
\VBF & \Hbb & \cite{HIGG-2019-04} \\
\ggF, \VBF, \WH + \ZH, \ttH + \tH & \HZZ & \cite{HIGG-2018-28} \\
\ggF, \VBF & \HWW & \cite{HIGG-2021-20} \\
\WH, \ZH & \HWW & \cite{HIGG-2017-14} \\
\ggF, \VBF, \WH + \ZH, \ttH + \tH & \Htautau & \cite{HIGG-2019-09} \\
\ggF + \ttH + \tH, \VBF + \WH + \ZH & $H \rightarrow \mu\mu$ & \cite{HIGG-2019-14} \\
Inclusive & $H \rightarrow Z\gamma$ & \cite{HIGG-2018-42} \\
\bottomrule
\end{tabular}
\label{tab:inputs}
\end{table}

An overview of the on-shell Higgs boson measurements is detailed in Ref.~\cite{HIGG-2021-23}.
Based on $pp$ collision data collected by the ATLAS experiment during LHC Run~2 at a centre-of-mass energy of $\sqrt{s}$~=~13~\TeV, corresponding to integrated luminosities ranging from 36.1~\ifb to 139~\ifb,
the \ttH cross-section is determined to be $0.37 \pm 0.12$~pb assuming SM Higgs boson decay branching ratios.
The measured value is in agreement with the SM prediction of $0.50\pm0.05$~pb~\cite{deFlorian:2016spz}.
The sensitivity is mainly contributed by the \Hyy~\cite{HIGG-2020-16}, \Hbb~\cite{HIGG-2020-23}, and multi-lepton~\cite{HIGG-2017-02} final states.
In the \Hyy analysis, a multi-class Boosted Decision Trees (BDT) classifier is first used to divide data into regions enriched with \ttH or Higgs bosons produced through other modes.
Then binary BDT classifiers are deployed to further discriminate the target signal from background processes.
The \ttH signal is extracted by fitting the diphoton invariant mass spectra in categories defined by the multi-class and binary BDT boundaries.
The background is estimated from data-driven interpolation in the data side-band.
The \Hbb analysis, on the other hand, first separates data into one-lepton and dilepton final states (electrons or muons), with the presence of $b$-tagged jets,
then uses either a BDT or a Deep Neural Network (DNN) algorithm to reconstruct the Higgs boson.
A classification BDT is used to further separate the \ttH signal from background processes.
The \ttH signal cross-section is extracted by fitting the classification BDT score distribution to data in most analysis categories.
The dominant background is from the \ttb process, which is modelled by Monte Carlo (MC) simulation with its normalisation determined from data.
Finally, the measurement from the multi-lepton final states~\cite{HIGG-2017-02}, which includes contributions from \HZZ, \HWW, and \Htautau decays, is removed from the work reported in this Letter due to its partial overlap with the \tttt data set~\cite{TOPQ-2021-08}.
After the removal, the overlap of data sets between the on-shell measurement and the \tttt\ measurement is negligible.

The measurement of \tttt\ production is detailed in Ref.~\cite{TOPQ-2021-08}. It uses $pp$ collision data at $\sqrt{s}$~=~13~\TeV from LHC Run~2, corresponding to an integrated luminosity of 140~\ifb. Events with multi-lepton final states are selected, specifically those with two leptons of the same electric charge or at least three leptons (electrons or muons). These events are also required to have at least two \(b\)-tagged jets. Major background processes include \ttW, \ttZ, and \ttH processes, with dedicated control regions defined to constrain the corresponding templates built from MC simulation. The rates of fake and non-prompt lepton background and charge mis-assignment background are also estimated and corrected using data-driven methods. A Graph Neural Network (GNN) classifier is used to further separate the \tttt\ signal from background processes. The \tttt\ cross-section is measured by fitting the GNN score distribution to data. The observed significance of the \tttt\ signal is 6.1 standard deviations ($\sigma$), with an expected significance of 4.3\,$\sigma$. The measured \tttt\ production cross-section is \( 22.5^{+6.6}_{-5.5} \)~fb, consistent with the SM prediction of $12.0 \pm 2.4$~fb~\cite{Frederix:2017wme, Bevilacqua:2012em, Jezo:2021smh} within 1.8\,$\sigma$.

\section{Combination framework}
\label{sec:comb}
The combined measurement uses the profile likelihood ratio and its asymptotic distribution to set upper limits on the total width of the Higgs boson~\cite{ATL-PHYS-PUB-2011-011,Cowan:2010js}.
The likelihood function of this combined measurement incorporates systematic uncertainties as constrained nuisance parameters and correlates those consistently defined between input measurements.
To incorporate the updated ATLAS integrated luminosity measurement~\cite{DAPR-2021-01} since the publication of Ref.~\cite{HIGG-2021-23}, the luminosity uncertainties in the on-shell measurement are revised and correlated with the \tttt measurement.
In addition, the uncertainties in the reconstruction and calibration of electrons, muons, jets, and missing transverse momentum, the uncertainty in the pile-up modelling by the MC simulation, and the theory uncertainties in \ttH, \tttt, and \ttZ MC modelling are all treated as correlated between the on-shell and \tttt measurements.
The impact from the potential correlation in other systematic uncertainties between the on-shell and \tttt measurements on the upper limit of \gammah is checked and found to be negligible.
The expected results corresponding to the SM prediction are evaluated based on an Asimov data set~\cite{Cowan:2010js}, which is generated with a set of nuisance parameters determined by a fit to data where the signal cross-sections are fixed to their SM values.

To parameterise the event rates of on-shell and off-shell Higgs boson production processes, the so-called \(\kappa\)-framework~\cite{LHCHiggsCrossSectionWorkingGroup:2013rie} is adopted.
The signal strength of any Higgs boson production and decay process \( i \rightarrow H \rightarrow f \) in the on-shell measurement, defined as the ratio of the total Higgs boson signal yield to its SM prediction, is parameterised as \( \kappai^2\kappaf^2/(\gammah/\gammah^{SM}) \) under the narrow-width approximation,
where \kappai\ and \kappaf\ are the coupling strength parameters modifying the production and decay rates, respectively,
and \( \gammah^{SM} \) is the SM value of the total width of the Higgs boson.
The strength parameters for the tree-level couplings to top quark (\kappat), bottom quark (\kappab), tau-lepton (\kappatau), muon (\kappamu), and weak bosons (\kappaz and \kappaw) are free parameters determined by the data.
The strength parameters for the three loop-induced effective couplings, namely, the Higgs-gluon coupling (\kappag), the Higgs-photon coupling (\kappay), and the Higgs-Z-photon coupling (\kappazy), are free parameters in the fit as well.
The coupling for the \( \ggZH \) process \( \kappaggzh \), on the other hand, is parameterised as a function of \( \kappaz \) and \( \kappat \). The current data set does not have the statistical power to constrain the \( \ggZH \) process. The \( \tttt \) process is a minor background in the \( \ttH \), \( \Hbb \) measurement, and its normalisation is fixed to the SM prediction within uncertainties. The constraint on the Higgs-top Yukawa coupling strength \( \kappat \) primarily comes from the \( \ttH \) production cross-section measurement. %
As a cross-check of this constraint, a fit is performed by parameterising  \gammah as a function of coupling strengths for SM particles based on the data used in the on-shell measurement.
The fit yields \( \kappat \) to be \( 0.86 \pm 0.13 \). This is 9\% lower than the value reported in Ref.~\cite{HIGG-2021-23} due to the removal of the  \( \ttH \) multi-lepton final states from the on-shell measurement. The uncertainty in \( \kappat \) also increases by 18\% when removing the \( \ttH \) multi-lepton final state.

For the \tttt\ measurement, the expected signal production rate in each bin of the GNN classifier discriminant is parameterised as a polynomial in \kappat, up to \({\cal{O}}(\kappat^4)\). The constant term arises from \tttt\ production mediated by gluons or \(Z/\gamma^{*}\), the \({\cal{O}}(\kappat^2)\) term arises from the interference between off-shell Higgs boson-mediated production and gluon- or \(Z/\gamma^{*}\)-mediated production, and the \({\cal{O}}(\kappat^4)\) term represents the contribution from off-shell Higgs boson-mediated production. There are no terms at odd orders of \kappat since there are always two Higgs-top Yukawa coupling vertices in the Higgs-mediated \tttt\ production diagrams. The polynomial is derived based on leading-order MC simulation samples~\cite{TOPQ-2021-08}. The \ttH\ process is included in the \tttt\ measurement as a background, and its normalisation is a free parameter determined from data. As discussed in Section~\ref{sec:results}, this treatment has a small impact on the determination of \gammah. Using this configuration, the \tttt\ measurement yields an observed (expected) 95\% CL upper limit on \(|\kappa_t|\) of 2.3 (1.9)~\cite{TOPQ-2021-08}. This observation is consistent with the SM at 1.8 standard deviations.

\section{Results}
\label{sec:results}

With \kappat and other Higgs boson coupling strength parameters as described in Section~\ref{sec:comb} profiled, the observed (expected) combined 95\% CL upper limit on \gammah\ is 450~\MeV\ (75~\MeV), corresponding to 110 (18) times the SM prediction.
The profile likelihood ratio as a function of \gammah\ is shown in Figure~\ref{fig:nll_1d}.
The observed value of the total width of the Higgs boson is $\gammah = 86^{+110}_{-49}$~\MeV, which is 2.0\,$\sigma$ away from the SM expectation. The tension with the SM prediction arises primarily from the 1.8\,$\sigma$ difference between data and SM prediction in the \tttt\ measurement.
Assuming the loop-induced \ggF, \Hyy, and \HZy\ rates can be modelled as a function of \kappat\ and other SM coupling strengths~\cite{LHCHiggsCrossSectionWorkingGroup:2013rie,HIGG-2021-23}, the observed (expected) 95\% CL upper limit on \gammah becomes 160~\MeV\ (55~\MeV) owing to the better constraint on \kappat contributed by these loop-induced processes. The deviation from the SM expectation in \gammah under this alternative scenario remains 2.0\,$\sigma$.

\begin{figure}[t]
\centering
\includegraphics[width=.48\textwidth]{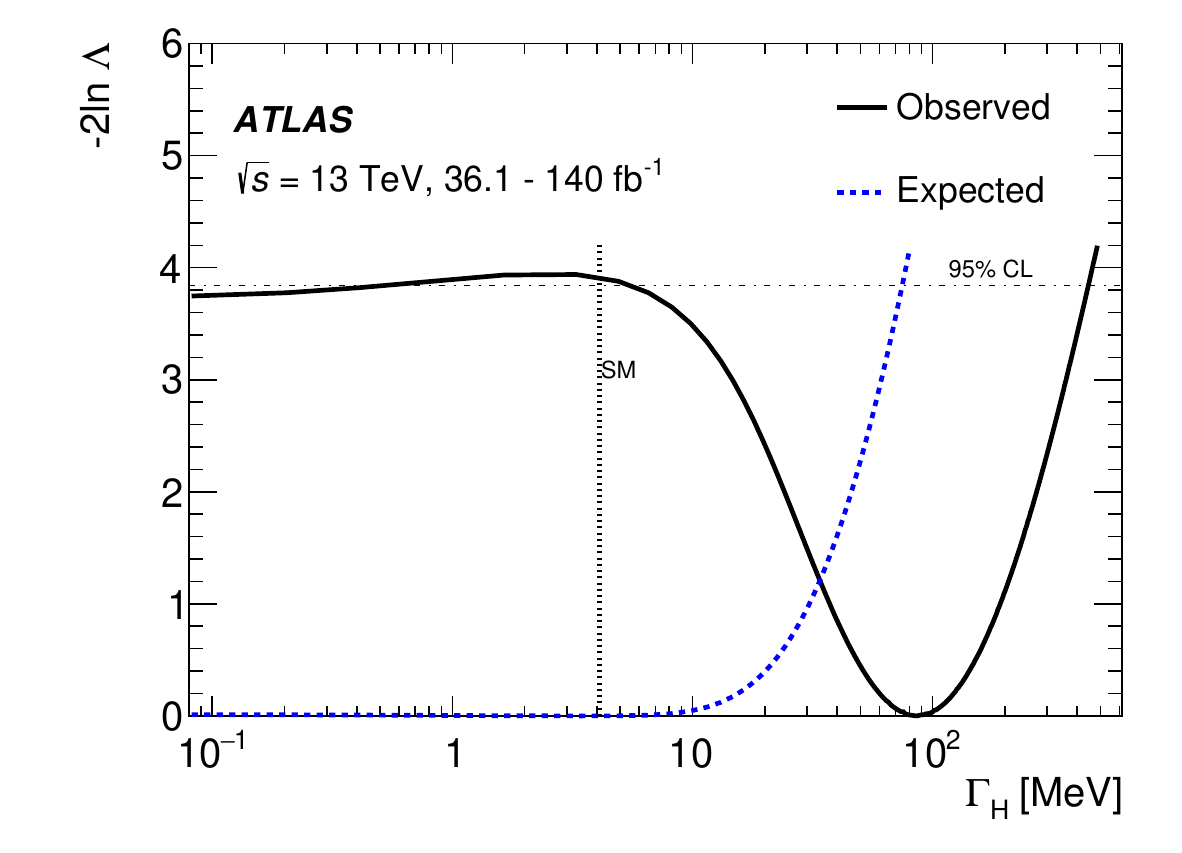}
\caption{The observed (expected) profile likelihood ratio, $-2\ln\Lambda$, as a function of \gammah is shown as a solid (dashed) line. The Higgs-top Yukawa coupling strength \kappat and other Higgs boson coupling strength parameters as described in Section~\ref{sec:comb} are profiled. The 95\% confidence interval is indicated by the intersections of the horizontal line with the $-2\ln\Lambda$ curves.
The vertical line indicates the SM prediction.}
\label{fig:nll_1d}
\end{figure}

The 68\% and 95\% CL contours in the \gammah\ and \kappat\ plane are shown in Figure~\ref{fig:nll_2d}. Because of the degeneracy among coupling strength parameters and \gammah in the on-shell measurement, the best-fit value of $\kappat =1.9$ is determined by the measurement of \tttt\ cross-section that is larger than the SM prediction. The best-fit value of \gammah is in turn scaled together with \kappat and other coupling strengths parameters along the flat direction of the likelihood function for the on-shell measurement.

\begin{figure}[t]
\centering
\includegraphics[width=.48\textwidth]{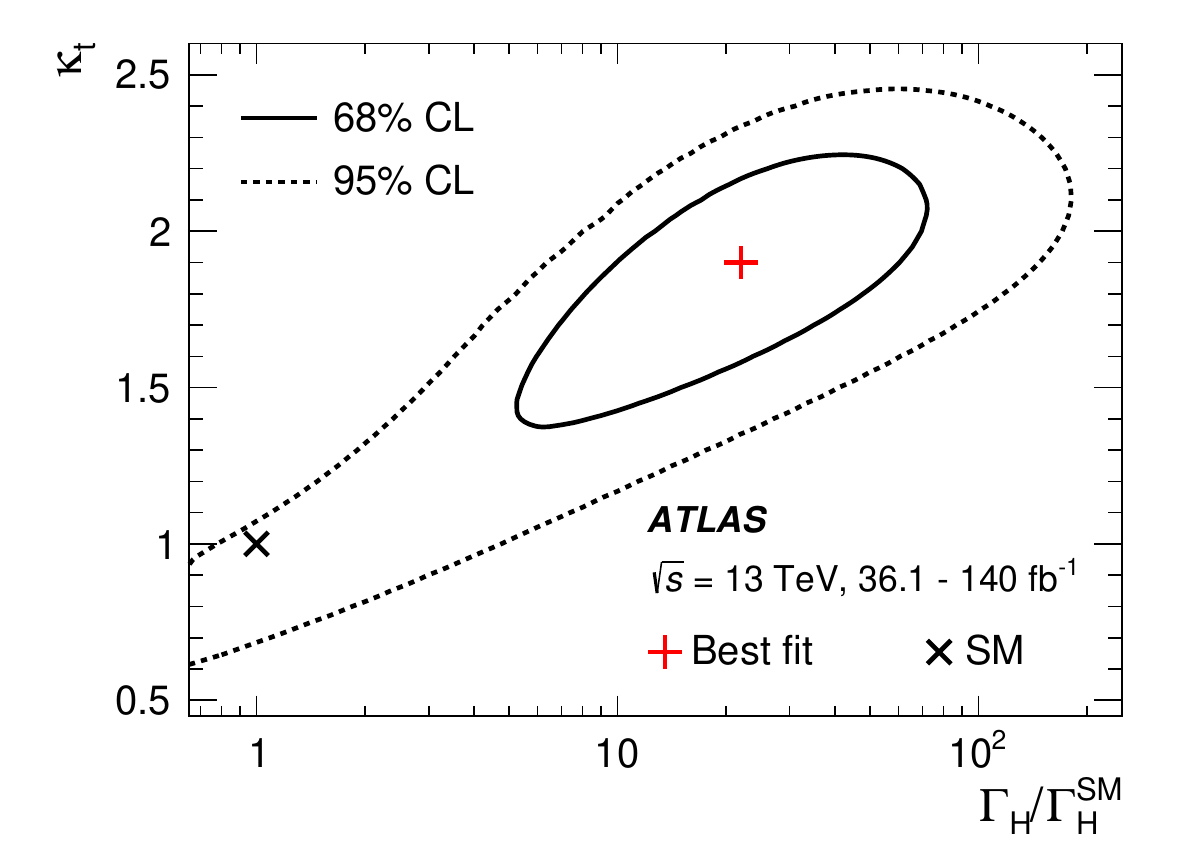}
\caption{The 68\% CL (solid line) and 95\% CL (dashed line) contours for a simultaneous measurement of \gammah\ normalised to the SM prediction and \kappat. The best-fit value and the SM prediction are also indicated in the figure.}
\label{fig:nll_2d}
\end{figure}

The Higgs boson decay contributions from the minor \ttH background in the \tttt\ measurement were not incorporated into the \(\kappa\)-framework based event rate parameterisation because the input measurement did not preserve information about the relative fractions of various Higgs decay final states in the signal region. However, this has minimal impacts on the interpretation. Up to a 6\% (2\%) decrease in the observed (expected) limit on \gammah\ is estimated when the \ttH\ background in the \tttt\ measurement is parameterised as functions of \gammah\ and \kappat\ in the combined fit, assuming either a 100\% \HWW\ or a 100\% \Htautau branching ratio. Note that the \HWW and \Htautau are the two leading Higgs boson decay channels that contribute to the multi-lepton final state targeted by the \tttt measurement. The assumption of full branching ratios is expected to provide a maximum estimate of the impact. Similarly, the input on-shell measurement did not consider the contribution from the \tttt\ process in constraining \kappat. When the \tttt\ normalisation in the \ttH, \Hbb\ measurement is parameterised as a function of \kappat~\cite{Cao:2016wib} in the combined fit, the observed (expected) limit on \gammah\ changes by 0.6\% (0.5\%).

The impact of various groups of systematic uncertainties in \gammah\ is evaluated by individually removing each group from the fit. This is achieved by fixing the associated nuisance parameters to their best-fit values and then quantifying the change in the upper limit on \gammah. As shown in Table~\ref{tab:systematics}, theory uncertainties have the most significant impact on the result. The largest impact on the expected limit, at 25\%, comes from the theory uncertainties in the \tttt\ process, which include missing higher-order QCD corrections evaluated from varying renormalisation and factorisation scales, MC generator choices, and the parton shower modelling. Details of the \tttt MC simulated sample are provided in Ref.~\cite{TOPQ-2021-08}.
Conversely, the largest impact on the observed limit, at 16\%, arises from the theory uncertainty in other physics processes, notably the \ttb background in the \ttH, \Hbb measurement. Due to a larger-than-SM best-fit value, the uncertainty in the \tttt\ cross-section measurement has a smaller observed impact on the \gammah\ measurement than in the expected case, owing to the non-linear mapping between the two measurements. Experimental uncertainties, such as those in jet energy calibration and flavour tagging, contribute only a 2\% impact on the limit. Removing all systematic uncertainties would decrease the observed (expected) upper limit to 280~\MeV\ (44~\MeV).

\begin{table}[t]
\centering
\caption{Impact of the main sources of systematic uncertainties in the expected and observed 95\% CL upper limit on the total width of the Higgs boson \gammah. The impact is quantified as the reduction of the upper limit when the corresponding systematic uncertainties are removed from consideration by fixing the associated nuisance parameters at the best-fit values. The impacts shown here are not a breakdown of uncertainties, and one does not expect that the quadratic sum of individual impact values will be the same as the total impact value.
}
\begin{tabular}{lcc}
\hline \hline
\multirow{2}{*}{Systematic uncertainty}                              & \multicolumn{2}{c}{Impact on 95\% CL upper limit on \gammah} \\
& Expected [\%] & Observed [\%] \\
\hline
Theory                                           & 37    & 33   \\
\hspace{3mm} \tttt\ production                       & 25    & 13   \\
\hspace{3mm} Higgs boson production/decay                  & 5     & 6    \\
\hspace{3mm} Other processes                        & 10    & 16   \\
\hline
Experimental                                     & 2     & 2     \\
\hspace{3mm} Jet flavour tagging                  & 2     & 1  \\
\hspace{3mm} Jet and missing transverse energy   & $<1$  & $<1$ \\
\hspace{3mm} Leptons and photons                 & $<1$  & $<1$ \\
\hspace{3mm} All other systematic uncertainties  & $<1$  & $<1$ \\
\hline \hline
\end{tabular}
\label{tab:systematics}
\end{table}

\FloatBarrier

\section{Conclusions}

In this Letter a constraint on the total width of the Higgs boson \gammah is obtained through a combination of measurements of on-shell Higgs boson production and \tttt production with contribution from the off-shell \Htoptop process. The input analyses are based on up to 140~\ifb of $pp$ collisions at a centre-of-mass energy of $\sqrt{s}$~=13~\TeV, collected with the ATLAS detector at the LHC\@. The study assumes that the Higgs-top Yukawa coupling strength \kappat remains the same between the on-shell and off-shell regimes. The resulting 95\% CL upper limit on \gammah is 450~\MeV, corresponding to 110 times the SM prediction, while the expected upper limit is 75~\MeV, corresponding to 18 times the SM prediction. Assuming that only SM particles contribute to the loop-induced gluon--gluon fusion, \Hyy, and \HZy\ processes, the observed (expected) upper limit decreases to 160~\MeV (55~\MeV). The tension between the data and the SM prediction is 2.0\,$\sigma$ in both scenarios, driven by the 1.8\,$\sigma$ difference between data and SM in the \tttt\ measurement. This result represents the first constraint on the total width of the Higgs boson using both on-shell and off-shell production processes involving the Higgs-top Yukawa coupling. It explores model assumptions distinct from those employed in similar studies based on diboson final states, thereby testing the robustness of our current understanding of the Higgs boson total width.

\FloatBarrier
\section*{Acknowledgements}
%

%

%
%

%
%

We thank CERN for the very successful operation of the LHC and its injectors, as well as the support staff at
CERN and at our institutions worldwide without whom ATLAS could not be operated efficiently.

The crucial computing support from all WLCG partners is acknowledged gratefully, in particular from CERN, the ATLAS Tier-1 facilities at TRIUMF/SFU (Canada), NDGF (Denmark, Norway, Sweden), CC-IN2P3 (France), KIT/GridKA (Germany), INFN-CNAF (Italy), NL-T1 (Netherlands), PIC (Spain), RAL (UK) and BNL (USA), the Tier-2 facilities worldwide and large non-WLCG resource providers. Major contributors of computing resources are listed in Ref.~\cite{ATL-SOFT-PUB-2023-001}.

We gratefully acknowledge the support of ANPCyT, Argentina; YerPhI, Armenia; ARC, Australia; BMWFW and FWF, Austria; ANAS, Azerbaijan; CNPq and FAPESP, Brazil; NSERC, NRC and CFI, Canada; CERN; ANID, Chile; CAS, MOST and NSFC, China; Minciencias, Colombia; MEYS CR, Czech Republic; DNRF and DNSRC, Denmark; IN2P3-CNRS and CEA-DRF/IRFU, France; SRNSFG, Georgia; BMBF, HGF and MPG, Germany; GSRI, Greece; RGC and Hong Kong SAR, China; ISF and Benoziyo Center, Israel; INFN, Italy; MEXT and JSPS, Japan; CNRST, Morocco; NWO, Netherlands; RCN, Norway; MNiSW, Poland; FCT, Portugal; MNE/IFA, Romania; MSTDI, Serbia; MSSR, Slovakia; ARIS and MVZI, Slovenia; DSI/NRF, South Africa; MICIU/AEI, Spain; SRC and Wallenberg Foundation, Sweden; SERI, SNSF and Cantons of Bern and Geneva, Switzerland; NSTC, Taipei; TENMAK, T\"urkiye; STFC/UKRI, United Kingdom; DOE and NSF, United States of America.

Individual groups and members have received support from BCKDF, CANARIE, CRC and DRAC, Canada; CERN-CZ, FORTE and PRIMUS, Czech Republic; COST, ERC, ERDF, Horizon 2020, ICSC-NextGenerationEU and Marie Sk{\l}odowska-Curie Actions, European Union; Investissements d'Avenir Labex, Investissements d'Avenir Idex and ANR, France; DFG and AvH Foundation, Germany; Herakleitos, Thales and Aristeia programmes co-financed by EU-ESF and the Greek NSRF, Greece; BSF-NSF and MINERVA, Israel; NCN and NAWA, Poland; La Caixa Banking Foundation, CERCA Programme Generalitat de Catalunya and PROMETEO and GenT Programmes Generalitat Valenciana, Spain; G\"{o}ran Gustafssons Stiftelse, Sweden; The Royal Society and Leverhulme Trust, United Kingdom.

In addition, individual members wish to acknowledge support from Armenia: Yerevan Physics Institute (FAPERJ); CERN: European Organization for Nuclear Research (CERN PJAS); Chile: Agencia Nacional de Investigaci\'on y Desarrollo (FONDECYT 1230812, FONDECYT 1230987, FONDECYT 1240864); China: Chinese Ministry of Science and Technology (MOST-2023YFA1605700, MOST-2023YFA1609300), National Natural Science Foundation of China (NSFC - 12175119, NSFC 12275265, NSFC-12075060); Czech Republic: Czech Science Foundation (GACR - 24-11373S), Ministry of Education Youth and Sports (FORTE CZ.02.01.01/00/22\_008/0004632), PRIMUS Research Programme (PRIMUS/21/SCI/017); EU: H2020 European Research Council (ERC - 101002463); European Union: European Research Council (ERC - 948254, ERC 101089007), Horizon 2020 Framework Programme (MUCCA - CHIST-ERA-19-XAI-00), European Union, Future Artificial Intelligence Research (FAIR-NextGenerationEU PE00000013), Italian Center for High Performance Computing, Big Data and Quantum Computing (ICSC, NextGenerationEU); France: Agence Nationale de la Recherche (ANR-20-CE31-0013, ANR-21-CE31-0013, ANR-21-CE31-0022, ANR-22-EDIR-0002), Investissements d'Avenir Labex (ANR-11-LABX-0012); Germany: Baden-Württemberg Stiftung (BW Stiftung-Postdoc Eliteprogramme), Deutsche Forschungsgemeinschaft (DFG - 469666862, DFG - CR 312/5-2); Italy: Istituto Nazionale di Fisica Nucleare (ICSC, NextGenerationEU), Ministero dell'Università e della Ricerca (PRIN - 20223N7F8K - PNRR M4.C2.1.1); Japan: Japan Society for the Promotion of Science (JSPS KAKENHI JP22H01227, JSPS KAKENHI JP22H04944, JSPS KAKENHI JP22KK0227, JSPS KAKENHI JP23KK0245); Netherlands: Netherlands Organisation for Scientific Research (NWO Veni 2020 - VI.Veni.202.179); Norway: Research Council of Norway (RCN-314472); Poland: Ministry of Science and Higher Education (IDUB AGH, POB8, D4 no 9722), Polish National Agency for Academic Exchange (PPN/PPO/2020/1/00002/U/00001), Polish National Science Centre (NCN 2021/42/E/ST2/00350, NCN OPUS nr 2022/47/B/ST2/03059, NCN UMO-2019/34/E/ST2/00393, NCN \& H2020 MSCA 945339, UMO-2020/37/B/ST2/01043, UMO-2021/40/C/ST2/00187, UMO-2022/47/O/ST2/00148, UMO-2023/49/B/ST2/04085, UMO-2023/51/B/ST2/00920); Slovenia: Slovenian Research Agency (ARIS grant J1-3010); Spain: Generalitat Valenciana (Artemisa, FEDER, IDIFEDER/2018/048), Ministry of Science and Innovation (MCIN \& NextGenEU PCI2022-135018-2, MICIN \& FEDER PID2021-125273NB, RYC2019-028510-I, RYC2020-030254-I, RYC2021-031273-I, RYC2022-038164-I), PROMETEO and GenT Programmes Generalitat Valenciana (CIDEGENT/2019/027); Sweden: Carl Trygger Foundation (Carl Trygger Foundation CTS 22:2312), Swedish Research Council (Swedish Research Council 2023-04654, VR 2018-00482, VR 2022-03845, VR 2022-04683, VR 2023-03403, VR grant 2021-03651), Knut and Alice Wallenberg Foundation (KAW 2018.0157, KAW 2018.0458, KAW 2019.0447, KAW 2022.0358); Switzerland: Swiss National Science Foundation (SNSF - PCEFP2\_194658); United Kingdom: Leverhulme Trust (Leverhulme Trust RPG-2020-004), Royal Society (NIF-R1-231091); United States of America: U.S. Department of Energy (ECA DE-AC02-76SF00515), Neubauer Family Foundation.

%
%


%
%
%
%
%
%

%
%
%
\printbibliography
\clearpage
 
\begin{flushleft}
\hypersetup{urlcolor=black}
{\Large The ATLAS Collaboration}

\bigskip

\AtlasOrcid[0000-0002-6665-4934]{G.~Aad}$^\textrm{\scriptsize 104}$,
\AtlasOrcid[0000-0001-7616-1554]{E.~Aakvaag}$^\textrm{\scriptsize 17}$,
\AtlasOrcid[0000-0002-5888-2734]{B.~Abbott}$^\textrm{\scriptsize 123}$,
\AtlasOrcid[0000-0002-0287-5869]{S.~Abdelhameed}$^\textrm{\scriptsize 119a}$,
\AtlasOrcid[0000-0002-1002-1652]{K.~Abeling}$^\textrm{\scriptsize 56}$,
\AtlasOrcid[0000-0001-5763-2760]{N.J.~Abicht}$^\textrm{\scriptsize 50}$,
\AtlasOrcid[0000-0002-8496-9294]{S.H.~Abidi}$^\textrm{\scriptsize 30}$,
\AtlasOrcid[0009-0003-6578-220X]{M.~Aboelela}$^\textrm{\scriptsize 45}$,
\AtlasOrcid[0000-0002-9987-2292]{A.~Aboulhorma}$^\textrm{\scriptsize 36e}$,
\AtlasOrcid[0000-0001-5329-6640]{H.~Abramowicz}$^\textrm{\scriptsize 155}$
% ...
\bigskip
\\

% ...
$^{17}$Department for Physics and Technology, University of Bergen, Bergen; Norway.\\
% ...
$^{30}$Physics Department, Brookhaven National Laboratory, Upton NY; United States of America.\\
% ...
$^{36}$$^{(a)}$Facult\'e des Sciences Ain Chock, Universit\'e Hassan II de Casablanca;$^{(b)}$Facult\'{e} des Sciences, Universit\'{e} Ibn-Tofail, K\'{e}nitra;$^{(c)}$Facult\'e des Sciences Semlalia, Universit\'e Cadi Ayyad, LPHEA-Marrakech;$^{(d)}$LPMR, Facult\'e des Sciences, Universit\'e Mohamed Premier, Oujda;$^{(e)}$Facult\'e des sciences, Universit\'e Mohammed V, Rabat;$^{(f)}$Institute of Applied Physics, Mohammed VI Polytechnic University, Ben Guerir; Morocco.\\
% ...
$^{45}$Physics Department, Southern Methodist University, Dallas TX; United States of America.\\
% ...
$^{50}$Fakult\"{a}t Physik , Technische Universit{\"a}t Dortmund, Dortmund; Germany.\\
% ...
$^{56}$II. Physikalisches Institut, Georg-August-Universit\"{a}t G\"ottingen, G\"ottingen; Germany.\\
% ...
$^{104}$CPPM, Aix-Marseille Universit\'e, CNRS/IN2P3, Marseille; France.\\
% ...
$^{119}$$^{(a)}$New York University Abu Dhabi, Abu Dhabi;$^{(b)}$United Arab Emirates University, Al Ain; United Arab Emirates.\\
% ...
$^{123}$Homer L. Dodge Department of Physics and Astronomy, University of Oklahoma, Norman OK; United States of America.\\
% ...
$^{155}$Raymond and Beverly Sackler School of Physics and Astronomy, Tel Aviv University, Tel Aviv; Israel.\\
% ...
$^{a}$ Also Affiliated with an institute covered by a cooperation agreement with CERN.\\
$^{b}$ Also at An-Najah National University, Nablus; Palestine.\\
$^{c}$ Also at Borough of Manhattan Community College, City University of New York, New York NY; United States of America.\\
$^{d}$ Also at Center for High Energy Physics, Peking University; China.\\
$^{e}$ Also at Center for Interdisciplinary Research and Innovation (CIRI-AUTH), Thessaloniki; Greece.\\
$^{f}$ Also at CERN, Geneva; Switzerland.\\
% ...

\end{flushleft}


%

%
%

%

%
%
%
%

\end{document}